\documentclass[english]{jetpl}
\twocolumn

\usepackage{amsmath,amssymb,epsfig,color,pstricks,bm,graphics,alltt}

\lat

\title{Electronic structure and possible pseudogap behavior in iron
based superconductors}

\rtitle{Electronic structure and possible pseudogap behavior}

\sodtitle{Electronic structure and possible pseudogap behavior in iron
based superconductors}

\author{E.\,Z.\,Kuchinskii\, 
M.\,V.\,Sadovskii\/\thanks{E-mail: sadovski@iep.uran.ru}}
\rauthor{E.\,Z.\,Kuchinskii, M.\,V.\,Sadovskii}

\sodauthor{Kuchinskii E.Z., Sadovskii M.V.}

\address{Institute for Electrophysics, Russian Academy of Sciences, Ural Branch,
620016 Ekaterinburg, Russia}

\dates{*}{*}

\abstract{

Starting from the simplified analytic model of electronic spectrum of
iron - pnictogen (chalcogen) high - temperature superconductors close to the
Fermi level, we discuss the influence of antiferromagneting (AFM)
scattering both for stoichiometric case and the region of possible
short -- range order AFM fluctuations in doped compounds. Qualitative picture of
the evolution of electronic spectrum and Fermi surfaces (FS) for different dopings is presented, with 
the aim of comparison with existing and future ARPES experiments. Both electron 
and hole dopings are considered and possible pseudogap behavior connected with 
partial FS ``destruction'' is demonstrated, explaining some recent experiments.

}


\begin{document}

\maketitle

Recent discovery of the new class of iron based high-temperature superconductors 
\cite{kamihara_08} stimulated intensive of experimental and 
theoretical efforts to understand its properties (see for the review Refs. 
\cite{UFN_90,Hoso_09}). Despite already the immense progress in understanding 
of these systems, the nature of superconducting pairing and anomalies in the
normal state are still under debate.

Clarification of the structure of electronic spectrum of new superconductors
is crucial for explanation of their physical properties. Accordingly, since
the first days, different groups have started the detailed band -- structure
calculations for all classes of these compounds, based primarily on different
realizations of general LDA approach.
These calculations were primarily performed for paramagnetic tetragonal FeAs 1111 
systems \cite{dolg,mazin,Xu1282,Nek1239}, for 122 \cite{Nek2630,Shein,Singh2643}, 
for 111 \cite{Singh2643,Nek1010,SheinIv} and $\alpha$-FeSe 
\cite{Singh4312}, followed by many similar works by other authors. 
In fact, all these calculations
demonstrated almost universal LDA band structure in relatively narrow energy
interval ($\pm 0.1 eV$) around the Fermi level, which is of relevance to superconductivity
\cite{UFN_90}. 

\begin{figure}
\includegraphics[clip=true,width=0.55\textwidth]{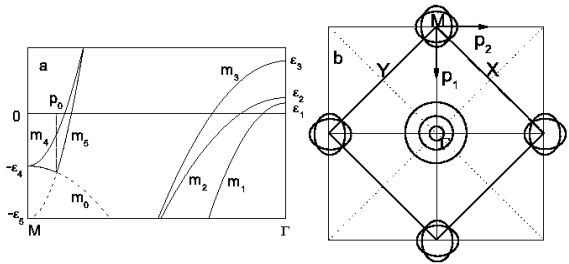}
\caption{Fig.1. Qualitative picture from of the band structure in M$\Gamma$ 
direction in the reduced Brillouin zone (a) and the relevant Fermi surfaces (b).}
\label{bare_sp}
\end{figure}

In this energy interval the electronic spectrum can be modelled analytically as follows.  
Three ``hole-like'' branches of the spectrum crossing the Fermi level near the  
$\Gamma$ point in the Brillouin zone (cf. Fig.\ref{bare_sp}a) can be taken 
isotropic and modelled by quadratic dispersion:

\begin{equation}
\varepsilon_i({\bf p})=\varepsilon_i-\frac{p^2}{2m_i}
\label{sp_hole}
\end{equation}
where $m_i,\ \varepsilon_i\ (i=1,2,3)$ can be easily determined from LDA calculations 
(e.g. for 122 system from the results of Ref. \cite{Nek2630}).

Two ``electron-like'' branches of the spectrum crossing the Fermi level near 
M$(\pi ,\pi )$ point of the reduced Brillouin zone are
anisotropic and produce two elliptic isoenergetic crossections at the Fermi
level (cf. Fig.\ref{bare_sp}b), one of which is extened in the direction 
M$\Gamma$, with the second one extended in the orthogonal direction. 
Let us count the momentum from the M point (i.e. replace ${\bf p-Q}\to{\bf p}$) 
and take one momentum ${\bf p}$ axis along M$\Gamma$ direction and other 
orthogonal to it (Fig.\ref{bare_sp}b). The relevant momentum projections $p_1$ 
and $p_2$ are connected with the usual $x,\ y$ projections as 
$p_1=\frac{p_y+p_x}{\sqrt{2}},\ p_2=\frac{p_y-p_x}{\sqrt{2}}$. Consider one
of the ellipses, e.g. those extended along the direction orthogonal to 
M$\Gamma$ direction. Electron dispersion along M$\Gamma$ can be
modelled by quadratic law $\varepsilon_{p1}( p)=\frac{p^2}{2m_4}-\varepsilon_4$. 
Dispersion along the orthogonal (to M$\Gamma$) direction is determined by
higher (in energy) branch of the spectrum, originating from the hybridization 
of two ``bare'' dispersions (cf. Fig. \ref{bare_sp}a), which we also assume
quadratic. Then, neglecting the small hybridization gap, 
we obtain $\varepsilon_{p2}( p)=Max(-\frac{p^2}{2m_0}-\varepsilon_4;\frac{p^2}{2m_5}-\varepsilon_5)$
Parameters $m_4,\varepsilon_4,m_5,\varepsilon_5,m_0$ can be taken from LDA data.
Thus, for anisotropic ``electron-like'' spectrum we can use the following
model:
\begin{equation}
\varepsilon_4({\bf p})=cos^2(\phi )\varepsilon_{p1}( p)+sin^2(\phi )\varepsilon_{p2}( p)
\label{sp_phi}
\end{equation}
where $p^2=p^2_1+p^2_2$, and $\phi$ is the polar angle with respect to 
$p_1$ axis. This model guarantees correct energy crossections in direction
M$\Gamma$ and orthogonal it, as well as isotropy of the spectrum in case of 
$\varepsilon_{p1}(p)=\varepsilon_{p2}(p)$. Energy dispersion for the second
``electron-like'' band $\varepsilon_5({\bf p})$ is also given by
Eq. (\ref{sp_phi}) with the obvious substitution $\phi\to\frac{\pi}{2}+\phi$.
Finally, we describe the ``electron-like'' bands in our model as: 
\begin{equation}
\varepsilon_4({\bf p})=\left\{\begin{array}{ll}
\frac{p^2_1}{2m_4}\!-\!\!\frac{p^2_2}{2m_0}\!-\!\!\varepsilon_4
&\mbox{\ for\ } p^2=p^2_1\!\!+\!\!p^2_2<p^2_0 \\
\frac{p^2_1}{2m_4}\!+\!\!\frac{p^2_2}{2m_5}
\!-\!\!\frac{p^2_1}{p^2}\varepsilon_4\!\!-\!\!\frac{p^2_2}{p^2}\varepsilon_5
&\mbox {\ for\ } p^2>p^2_0
\end{array} \right.
\label{sp_el}
\end{equation}
where $p^2_0=2(\varepsilon_5-\varepsilon_4)/(\frac{1}{m_5}+\frac{1}{m_0})$ 
is the square of the momentum at the crossing of two ``bare'' hybridizing
bands.

The qualitative picture of electronic spectrum and Fermi surfaces is shown
in Fig. \ref{bare_sp}.
Essentially this kind of electronic spectra and Fermi surfaces in new 
superconductors were qualitatively confirmed by angle resolved
photoemission spectroscopy (ARPES), starting with the early works 
\cite{Yang2627,Liu3453,Liu4806,Zhao0398,Ding0419,Zab2454,Evt4455,Sato3047}, 
followed by many further studies by the same and other authors. 
Most of these experiments were performed on single crystals of 122 systems,
while for other compounds good quality single crystals are up to now just
unavailable. Though in qualitative agreement with the results of LDA
calculations, these experiments show rather different results concerning 
finer details, such as the precise number of ``hole-like'' FS cylinders around 
the $\Gamma$ point, as well as the topology of ``electron-like'' cylinders 
around the M -- point.

In general LDA calculations underestimate the role of electronic correlations.
ARPES experiments show that these systems apparently belong to the class of
intermediately correlated systems, with correlation induced band narrowing by
the factor of two \cite{Liu4806}. This is confirmed by some of LDA+DMFT 
calculations \cite{AnIzV}, though theoretical situation here remains rather 
controversial. In the following we take correlations into account by simple 
rescaling of the energy by the factor of two as compared with LDA 
\cite{Liu4806}.

Undoped FeAs compounds are antiferromagnetically ordered with AFM vector
${\bf Q}=(0,\pi)$ in extended Brillouin zone, corresonding to ${\bf Q}=(\pi,\pi)$
in the reduced zone \cite{UFN_90,Hoso_09}. Electron or hole doping suppresses 
AFM ordering and induces superconductivity, similar to the well known situation 
in cuprates. Recent neutron scattering experiments \cite{Inos,Dial} clearly
show that in the substantial part of the phase diagram of FeAs systems in
normal paramagnetic state rather strong fluctuations of AFM short-range order
persist, as predicted e.g. by the model of ``nearly antiferromagnetic Fermi
liquid'' \cite{NAFL,Mont1,Mont2}. These fluctuations can, in principle, induce 
the pseudogap behavior in electronic spectrum, similar to that observed in
cuprates \cite{pseudogap}. 

\begin{figure}
\includegraphics[clip=true,width=0.45\textwidth]{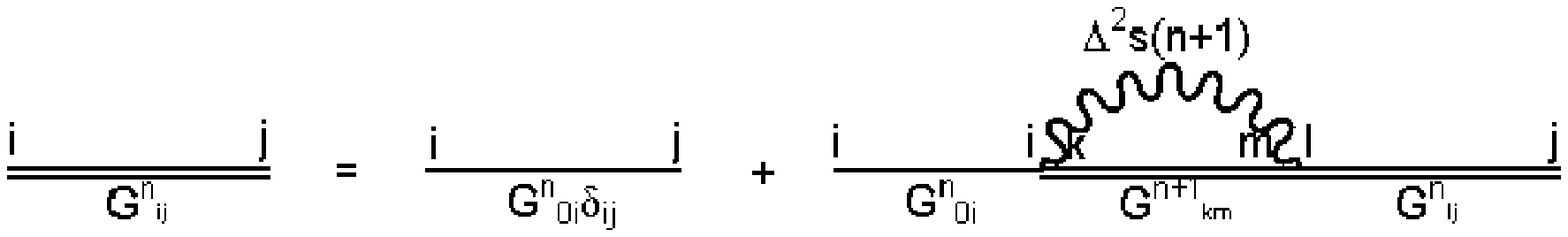}
\caption{Fig.2. Recurrence ``Dyson equation'' for the Green's function.}
\label{diagg}
\end{figure}

Effective interaction of electrons with AFM spin fluctuations is determined in this
model by dynamic spin susceptibility characterized by the maximum at scattering vectors
close to AFM vector ${\bf Q}=(\pi,\pi)$, which we assume here to be the same for electron
from different bands and for interband scattering. Limiting ourselves to high enough
temperatures we can neglect the dynamics of AFM fluctuations and consider them Gaussian 
\cite{pseudogap}. The Green's function for electrons moving in the ``quenched'' Gaussian 
random field of these fluctuations can be represented by recurrence ``Dyson equation'' shown 
in Fig.\ref{diagg}, which is the direct multiple bands generalization of the summation
procedure, proposed and actively used in Refs. \cite{MS79,Diagr,Sch,KS}, taking into account 
{\it all} Feynman diagrams for electron scattering in such random field. 

Analytically, this ``Dyson equation'' can be written as:
\begin{equation}
G_{ij}^{n}=G_{0i}^{n}\delta_{ij}+
G_{0i}^{n}\Delta^2s(n+1)\sum_{km}G_{km}^{n+1}\sum_{l}G_{lj}^{n}
\label{G_ij}
\end{equation}
where $i,j$ represent band indices, $\Delta$ characterizes the AFM pseudogap width
(of the order of AFM band splitting),
\begin{equation}
G^n_{0i}(E{\bf p})=\frac{1}{E -\varepsilon^n_i({\bf p})+inv^n_i\kappa}
\label{G0}
\end{equation}
$\kappa=\xi^{-1}$ is an inverse correlation length of AFM short-range order fluctuations, 
$\varepsilon^n_i({\bf p})=\varepsilon_i({\bf p+Q})$ and $v^n_i=|v^x_i({\bf p+Q})|+|v^y_i({\bf p+Q})|$ 
for odd $n$, while  
$\varepsilon^n_i({\bf p})=\varepsilon_i({\bf p})$ and $v^n_i=|v^x_i({\bf p})|+|v^y_i({\bf p})|$ 
for even $n$. Velocity projections $v^x_i({\bf p})$ and $v^y_i({\bf p})$ are determined by the momentum
derivatives of electronic dispersion in the $i$-th band $\varepsilon_i({\bf p})$. 
Combinatorial factor $s(n)$ for the case of Heisenberg AFM fluctuations
(spin-fermion model of Ref. \cite{Sch}) is given by: 
\begin{equation} 
s(n)=\left\{\begin{array}{cc}
\frac{n+2}{3} & \mbox{for odd $n$} \\
\frac{n}{3} & \mbox{for even $n$}.
\end{array} \right.
\label{vspin}
\end{equation}
The physical Green's function corresponds to $n=0$. 
Then, after some simple manipulations we may show that
\begin{equation}
G_{ij}(E{\bf p})=G^0_{0i}(E{\bf p})\delta_{ij}+
\frac{G^0_{0i}(E{\bf p})G^0_{0j}(E{\bf p})\Sigma (E{\bf p})}
{1-G^0_{0}(E{\bf p})\Sigma (E{\bf p})}
\label{Gij}
\end{equation}
where the physical self-energy
\begin{equation}
\Sigma (E{\bf p})=\Sigma^{n=1}(E{\bf p})
\label{Sigma_ph}
\end{equation}
is determined from the recurrence procedure (continued fraction representation):
\begin{equation}
\Sigma^n(E{\bf p})=\frac{\Delta^2s(n)}
{(G^n_{0}(E{\bf p}))^{-1}-\Sigma^{n+1}(E{\bf p})}
\label{G_rec}
\end{equation}
where $G^n_{0}(E{\bf p})=\sum_{j}G^n_{0j}(E{\bf p})$.
As a byproduct of these general equations we can easily analyze the electronic
spectrum in the case of AFM long-range order, truncating the continuous
fraction in Eq. (\ref{G_rec}) at $n=1$ and taking the limit of $\kappa\to 0$. 
The spectral density and density of states are obviously given by:
\begin{equation}
A(E{\bf p})=-\frac{1}{\pi} Im Sp G^{R}_{ii}(E{\bf p});\ 
N(E)=\sum_{\bf p} A(E,{\bf p})
\label{SPDOS}
\end{equation}

We performed calculations for a variety of parameters of the model, using for the 
spectrum LDA data for 122 from Ref \cite{Nek2630}, scaled by factor of two to account for 
correlations. Below we present results for $\Delta=$ 50 meV, which is in rough
agreement with the estimates of AFM band splitting from ARPES data 
\cite{Yan,Zhng} and neutron scattering \cite{Knl} (varying in the interval
50-100 meV), correlation length of AFM fluctuations $\xi=10a$ ($a$ - lattice 
spacing), also in rough agrrement with netron scattering data \cite{Inos,Dial}.
In the following, all momenta are given in units of inverse lattice spacing, 
energies in eV. To make the results comparable with ARPES
experiments we have also introduced effective widening to simulate finite
energy resolution of ARPES replacing $E\to E + i\gamma$ and taking
$\gamma$=10 meV (corresponding to best ARPES resolution).

\begin{figure}
\includegraphics[clip=true,width=0.75\textwidth]{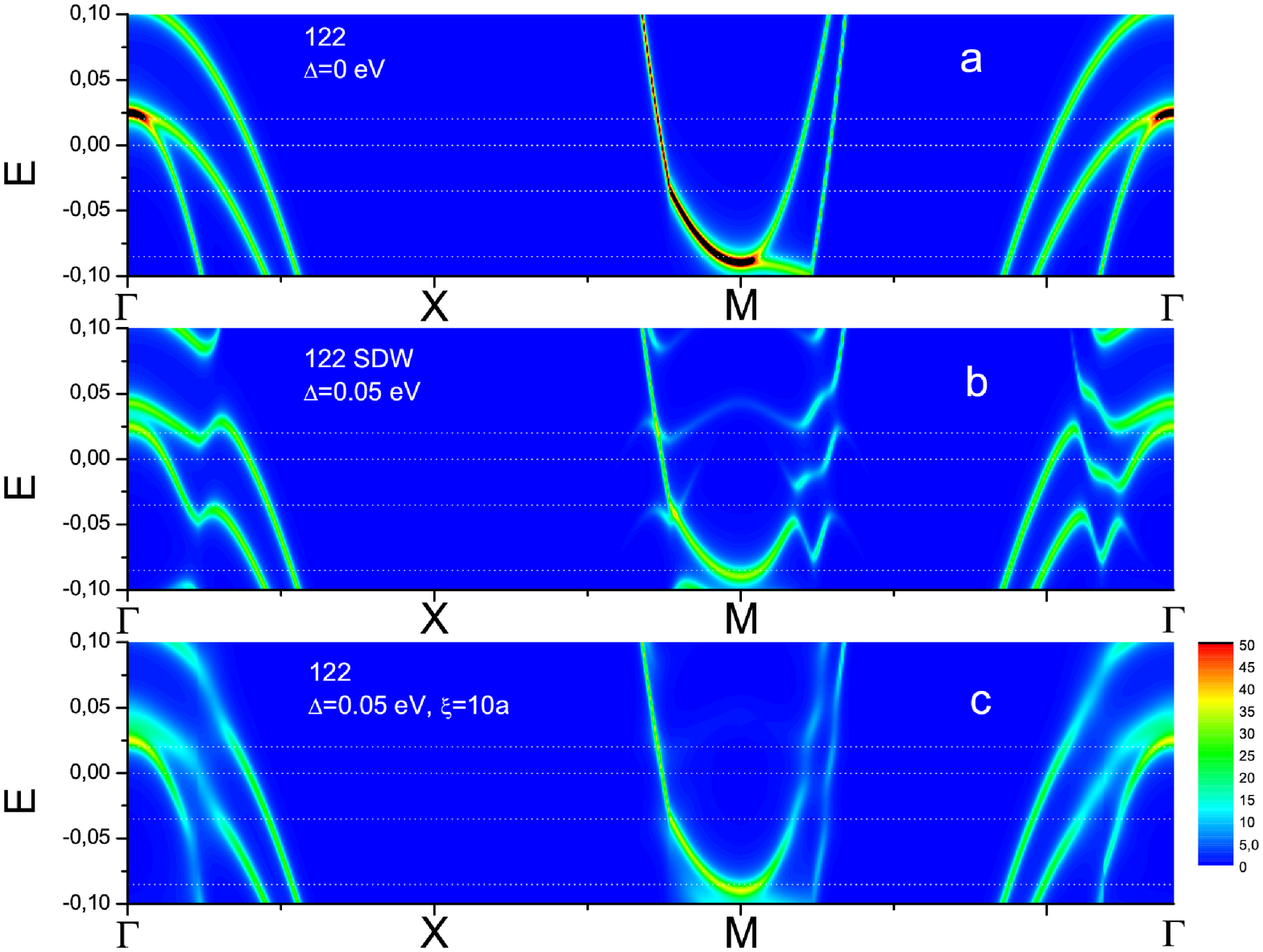}
\caption{Fig.3. Energy bands. Upper panel (a) -- ``bare'' (scaled LDA) bands in
paramagnetic state in the absense of AFM fluctuations. Panel  (b) -- 
AFM long-range ordered state with $\Delta=$0.05 eV. Panel (c) -- bands in the
pseudogap state induced by AFM short-range order fluctuations with 
$\xi =10a$ and $\Delta =$0.05 eV. All bands are shown with finite ``experimental''
resolution $\gamma=$0.01 eV. Dotted lines show Fermi levels for different
dopings used in our calculations of Fermi surfaces below.}
\label{bands}
\end{figure}

In Fig. \ref{bands} we show ``ARPES'' energy bands of 122 system, revealed by the maps
of spectral density, along main symmetry directions, starting from the case
of normal (paramagnetic) LDA bands, via AFM long-range ordered state, to
``pseudogapped'' state, characterized by electrons scattered by short-range
ordered AFM fluctuations -- AFM band splittings transforming 
to pseudogaps due to AFM short-range order.

\begin{figure}
\includegraphics[clip=true,width=0.7\textwidth]{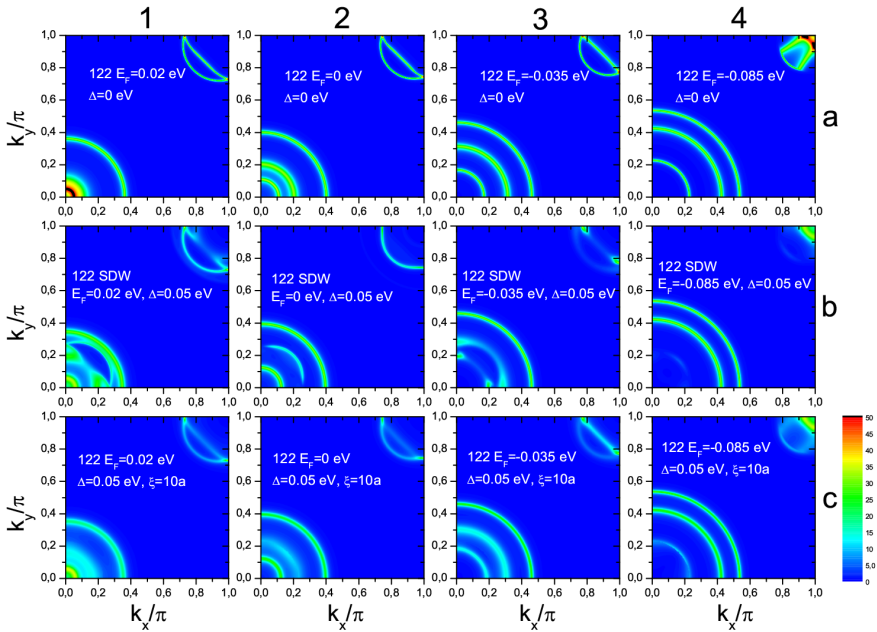}
\caption{Fig.4. ``ARPES'' Fermi surfaces at different doping levels shown by
dotted lines in Fig.\ref{bands}:  
Column 1 -- electron doping with $E_F=$0.02 eV, 
2 -- undoped system with $E_F=$0, 
3 -- hole doping with $E_F=$-0.035 eV (hole underdoped system), 
4 -- optimal hole doping with $E_F=$-0.085 eV. 
Upper panel (a) -- ``bare'' FS in paramagnetic state in the absense of  
AFM fluctuations. Panel (b) -- AFM phase with $\Delta=$0.05 eV. Panel (c) -- 
pseudogap state with $\xi =$10$a$ and $\Delta=$0.05 eV.}
\label{FSurf}
\end{figure}

In Fig. \ref{FSurf} we show spectral density maps at the Fermi level for different 
dopings -- from slightly electron doped, via undoped, to hole underdoped and
optimally hole doped case. These maps essentially produce ``ARPES'' Fermi
surfaces of 122 system at different dopings. In fact, the system always remains
metallic in a sense that at every doping we observe ``open'' Fermi surfaces,
though we also can see rather complicated series of Fermi surface transformations,
with some cylinders being almost ``destructed'' (damped) either by AFM
long-range order, or by short-range order AFM fluctuations. Of these maps, we
identify the last one in the third row (4c) as corresponding more or less to  
optimally hole doped case in satisfactory agreement with ARPES data e.g. from
Refs. \cite{Liu4806,Ding0419,Evt4455,Xu}, while the third one in the same row (3c)
apparently well corresponds to the hole underdoped case studied in Ref. \cite{Xu},
demonstrating the inner hole cylinder rather damped by pseudogap fluctuations
with characteristic wave vector of the order of AFM vector ${\bf Q}$. 
Significant pseudogap forms in the (partial) density of states on precisely
this cylinder, in agreement with Ref. \cite{Xu}.           
In general, the available ARPES data suffer from rather bad resolution, so that
pseudogap fluctuations can significantly complicate observation of all FS
cylinders and much work is needed to reveal possible complicated picture of FS
transformations, illustrated in Fig. \ref{FSurf}. It should be taken into
account that pictures shown in the second row (b) of Fig. \ref{FSurf} are sensible only
within the part of the phase diagram with AFM long-range order, while the
third row (c) applies to paramagnetic region, where superconductivity appears at
lower temperatures.                      

Our calculations show, that the pseudogap forms only in (partial) densities of 
states, corresponding to those cylinders strongly affected by short-range AFM
fluctuations, and this is not, in general, ``pinned'' at the Fermi level.
Pseudogap in the total density of states is always rather weak, and the problem
remains, whether it is sufficient to explain claims for the pseudogap
behavior observed in some NMR experiments \cite{UFN_90,Hoso_09}. 

This work is partly supported by RFBR grant 08-02-00021 and Programs of 
Fundamental Research of the Russian Academy
of Sciences (RAS) ``Quantum physics of condensed matter'' (09-$\Pi$-2-1009) and 
of the Physics Division of RAS  ``Strongly correlated electrons in solid states'' 
(09-T-2-1011).

\newpage

\onecolumn

\end{document}